\documentclass[prb,aps,floatfix,twocolumn,amsmath,amssymb,showpacs]{revtex4}
\newcommand{\ket}[1]{\ensuremath{\left|{#1}\right\rangle}}
\newcommand{\bra}[1]{\ensuremath{\left\langle{#1}\right|}}

\usepackage{epsfig}
\newcommand{\be}{\begin{equation}}
\newcommand{\ee}{\end{equation}}
\newcommand{\bea}{\begin{eqnarray}}
\newcommand{\eea}{\end{eqnarray}}
\def\nn{\nonumber\\}

\begin{document}
\DeclareGraphicsExtensions{.eps}
%%%%%%%%%%%%%%%%%%%%%%%%%%%%%%%%%%%%%%%%%%%%%%%%%%%%%%%%%%%%%%%%
\title{Finite Temperature Dynamical Structure Factor
of Alternating Heisenberg Chains}
%%%%%%%%%%%%%%%%%%%%%%%%%%%%%%%%%%%%%%%%%%%%%%%%%%%%%%%%%%%%%%%%
\date{\today}
\author{A. J. A. James$^1$}
\author{F. H. L. Essler$^1$}
\author{R. M. Konik$^2$}
\affiliation{$^1$Rudolf Peierls Centre for Theoretical Physics, University of Oxford, Oxford OX1 3NP, UK \\
$^2$Brookhaven National Laboratory, Upton, NY 11973 USA}
\begin{abstract}
We develop a low-temperature expansion for the finite temperature
dynamical structure factor of the spin half Heisenberg chain with
alternating nearest neighbour exchange in the limit of strong
alternation of the exchange constants. We determine both the
broadening of the low lying triplet lines and the contribution of the
thermally activated intraband scattering. 
\end{abstract}

\maketitle
%%%%%%%%%%%%%%%%%%%%%%%%%%%%%%%%%%%%%%%%%%%
\section{Introduction}
%%%%%%%%%%%%%%%%%%%%%%%%%%%%%%%%%%%%%%%%%%%
Much is now known about the physics of quasi-one dimensional
Heisenberg anti-ferromagnetic chain materials at zero
temperature. This understanding has benefited from powerful analytical
techniques (see e.g. \cite{Affleck89b,Bougourzi97,review,Caux} and
references therein) as well as highly accurate experiments 
performed at temperatures much smaller than the relevant exchange
constants\cite{exp}. As such the regime in which these materials are understood
is dominated by quantum fluctuations. In contrast, far less is known
about their finite temperature behaviour where there is an interplay
of quantum and thermal fluctuations \cite{sachdevbook,young,damle,finiteT,mikeska,essrmk}. 

Though the chemistry of these materials can be exceedingly
complicated, a wide range of their properties, up to energy scales set
by their exchange constants, are well described by simple lattice
models of the form  
\begin{equation}
\label{ei}
H = \sum_{i,j} J_{ij} {\bf S}_{i} \cdot {\bf S}_{j}.
\end{equation}
Here $J_{ij}$ is the exchange constant between spins at sites $i$ and $j$.

Perhaps the most encompassing probe of the properties of spin chains are
inelastic neutron scattering experiments. Such experiments provide 
detailed information on the chains' excitations\cite{igor}.
In theoretical terms, these experiments specifically yield the
spin dynamical structure factor
\begin{align}
\label{eii}
\begin{split}
S^{\alpha\gamma}(\omega, Q)&=-\frac{1}{\pi}\frac{1}{1-e^{-\beta\omega}}
{\rm Im}\left[
\chi^{\alpha\gamma}(\omega, Q)\right] ,\\
\chi^{\alpha\gamma}(\omega, Q)&=-\int_0^\beta d\tau\ e^{i\omega_n\tau}\\
& \quad\frac{1}{N}\sum_{l,l'}e^{-iQ(l-l')} \langle S^\alpha_l(\tau)\
S^\gamma_{l'}\rangle\biggr|_{\omega_n\rightarrow \eta-i\omega}.
\end{split}
\end{align}
Here $\beta=1/k_BT$ and $\alpha,\gamma=x,y,z$ and the brackets imply a thermal expectation. We have used the Matsubara formalism, with imaginary times $\tau$ and frequencies $\omega_n$. The dynamical
susceptibilities $\chi^{\alpha\gamma}(\omega, Q)$ take their 
simplest form at zero temperature, where they provide direct
information on the energies and lifetimes of the spin excitations.
At finite temperature the $\chi^{\alpha\gamma} (\omega, Q)$ become more
complex functions, determined now by a competition between thermal 
and quantum effects.  
The value of the exchange coupling, $J$, determines the extent of the
temperature's role. If $J \gg T$, there exists an appreciable range of
energies, $\omega$, over which inelastic neutron scattering
experiments effectively probe the zero temperature form of
$\chi^{\alpha\gamma} (\omega, Q)$. However if $J \sim T$ or one is
interested in energies, $\omega$, on the order of $T$, the effects
of temperature must be taken into account when calculating $\chi^{\alpha\gamma} (\omega, Q)$. 

In this article, we are interested in calculating the dynamical
structure factor at low but finite temperature for a class of dimerized 
spin-1/2 chain materials. These materials are well described
by the Hamiltonian
\begin{equation}\label{eiii}
H_{\rm dimer} = \sum_{a=0}^{N/2-1}\left( J \mathbf{S}_{2a}
\cdot  \mathbf{S}_{2a+1} +J' \mathbf{S}_{2a+1} \cdot
\mathbf{S}_{2a+2} \right). 
\end{equation}
Here $J$ is the exchange coupling through which
pairs of neighbouring $S=1/2$ spins form singlet dimers while $J' = \alpha
J$ gives the strength of the interdimer interaction. We will be
interested in the case $\alpha \ll 1$. The lowest lying excitations
of this model, which we will refer to as magnons, are characterized by
a gap, $\Delta$, which at zeroth order in $\alpha$ represents  the
cost of breaking one dimer.  With small but finite $\alpha$, the
magnons disperse according to 
$$
\epsilon_k = J - \frac{\alpha J}{2} \cos (k d),
$$
where $d$ is the interdimer distance.
A particular realization of this material is 
${\rm Cu(NO_3)_2\cdot 2.5H_2O}$
where $J = 5.22 K$ and $\alpha = 0.27$\cite{tennant,xu1}.
While this material has been studied by inelastic neutron scattering at
$\rm 300mK$\cite{tennant}, a temperature far smaller than $\Delta =
4.4K$, more recent experiments have been performed\cite{tennant1} at
temperatures on the same order as the gap where thermal effects  
on the form of $\chi^{\alpha\gamma} (\omega, q)$ cannot be ignored.
Previously much theoretical work has concentrated on the zero
temperature limit, establishing various properties of the spectrum of
(\ref{eiii}), including multi-particle continua and bound
states\cite{tennant2,uhrig,brooksharris,zheng,bouzerar,trebst,schmidt1,collins} as well as spectral weights\cite{schmidt2,hamer1} and the dynamical structure factor\cite{essdsf,hamer1,schmidt1}. However at finite temperature, the most
pertinent features to explore are the broadening of the single
particle modes, due to interaction with the thermally populated
background, and the low frequency response arising from intraband
transitions. 

By virtue of the spin rotational invariance of the Hamiltonian
(\ref{eiii}) all off-diagonal elements of the dynamical susceptibility
vanish and the three diagonal elements are the same, i.e.
$\chi^{xx}(\omega,Q)=\chi^{yy}(\omega,Q)=\chi^{zz}(\omega,Q)$. In what
follows we will therefore only consider $\chi^{zz}$.

Accounting for the effects of temperature in the dynamical
susceptibilities makes this problem particularly
challenging. To see why we consider a Lehmann expansion of the spin
response function in terms of the eigenstates of the model,
$\{|l\rangle\}$.  
Defining $C(\tau, x) = \langle S^z_j(\tau) S^z_k\rangle$
where $x=R_j-R_k$, this expansion takes the form, 
\begin{equation}\label{eiv}
C(\tau, x) = 
\frac{1}{Z}\sum_{l,m}e^{-\beta E_l}
\langle l | S^z_j(\tau)|m\rangle
\langle m | S^z_k|l\rangle.
\end{equation}
The double sum in this representation for $C(\tau,x)$ 
arises on the one hand from the Boltzmann sum, $\sum_l e^{-\beta
  E_l}$, where $E_l$ is the energy of eigenstate $|l\rangle$, and on
the other hand from an insertion of a resolution of the identity
between the two operators, $S^z_j(\tau)$ and $S_k^z$. 
This expansion renders the task of finding $C(\tau, x)$
into a matter of computing individual matrix elements 
$\langle l | S_j^z(\tau)|m\rangle$.  At zero temperature, this
computation is simplified on two counts:\cite{review} (i) the sums over
eigenstates in Eqn. (\ref{eiv}) reduce to a single sum; and (ii) the
matrix elements needed are of a single type, namely those connecting the ground
with various excited states. At finite temperature however, we must
deal both with the double sum in Eqn. (\ref{eiv}) and the matrix
elements in their full generality. 

To make this task tractable, we exploit the fact that the spin chain
material is gapped. On a qualitative level the excitations can be divided
according to the number, $n$, of magnons they contain.  The energy of
an excitation with $n$ magnons is then at least $n\Delta$.  This
notion is imprecise because magnon number is not a good quantum number,
nonetheless at small $\alpha$ it can serve as a rough guide to the
excitations' energies.  In turn, provided the temperature does not
exceed the gap, $\Delta$, the contribution to the sum in Eqn
(\ref{eiv}) of excitations containing large numbers of magnons will be
exponentially suppressed by the Boltzmann factor, $e^{-\beta E_l}$.
In such a case, we thus need only to consider excitations in the
Boltzmann sum, $\sum_l e^{-\beta E_l}$, involving only a few magnons.
Concomitantly, provided we are interested in determining ${\rm
  Im} \chi^{zz}(\omega,Q)$ at energies not far
in excess of the gap, we can similarly restrict the sum in
Eqn.(\ref{eiv}), $\sum_m$, arising from the resolution of the identity. 

The evaluation of $\chi^{zz}(\omega, Q)$ is, however, more
delicate than the above implies. When evaluating the leading terms in
the Fourier transform of the spectral representation (\ref{eiv}), one
finds divergences when the frequency approaches the magnon
dispersion. Such divergences are expected, since the spectral sum
still contains the $T=0$ result, which is a delta-function at the
position of the single-magnon dispersion. On physical grounds the
single-magnon line is expected to broaden at $T>0$. Analytically this is achieved
by carrying out a resummation on the divergences of the
higher order terms in the expansion. Specifically, the sum in Eqn. (\ref{eiv}) can be reorganized according to a Dyson-like
equation,\cite{essrmk} where we write $\chi^{zz}(\omega, Q)$ in the
form, 
\begin{equation}\label{ev}
\chi^{zz}(\omega, Q) = 
\frac{D(\omega, Q)}{1 - D(\omega, Q)\Sigma (\omega, Q)}.
\end{equation}
Here $D(\omega, q)$ can be thought of as the propagator for
non-interacting magnons and $\Sigma(\omega, q)$ 
is the magnon self-energy. The key is to match the perturbative
expansion of (\ref{ev})
\be
\label{evi}
\chi^{zz}(\omega, Q) = D(\omega,Q)+D^2(\omega,Q)\Sigma(\omega,Q)+\ldots
\ee
to the spectral representation of $\chi^{zz}(\omega,Q)$, which is
given in terms of the Fourier transform of (\ref{eiv}). In this way we
obtain a controlled low temperature expansion of the self-energy,
$\Sigma(\omega, q)$, in lieu of $\chi^{zz}(\omega, Q)$.

This approach has been used successfully in the study of finite
temperature dynamical correlation functions in gapped one dimensional
quantum antiferromagnets with continuum integrable field theoretic
representations \cite{essrmk}. There the matrix elements, $\langle
l|S^z_j(0)|m \rangle$, were computed exactly via analyticity constraints
coming from integrability.\cite{smirnov,review}  However, the
model of the dimerized spin chain, Eqn. (\ref{eiii}), is not exactly
solvable. But because $\alpha$ is small, we can compute the necessary
matrix elements perturbatively in $\alpha$.

The $T>0$ dynamical susceptibility has been studied previously using exact diagonalization
of finite length chains.\cite{mikeska}  
We believe our approach provides a useful complement to this work.
The numerical approach yields results for all $\alpha$ and is not restricted to small temperatures. However, the
system size that can be studied is quite small.
We, on the other hand, 
must proceed perturbatively in $\alpha$ and are restricted to low
temperatures, but our calculations do not suffer from 
finite-size effects.  Moreover the nature of low-lying excitations 
is more apparent and we can identify the specific processes that
give rise to the various finite-temperature effects in the structure
factor. 

With this in mind, a specific feature that we focus upon in our
analysis is the presence of temperature induced neutron scattering
intensity at low frequencies much smaller than the zero temperature
gap. The origin of this intensity is intraband scattering. The
analogous phenomenon in Ising-like antiferromagnetic spin chains was
first pointed out by J. Villain \cite{villain} 
and was first observed in the anisotropic spin chain material
$\rm CsCoBr_3$.\cite{nagler,nagler1}  For Ising antiferromagnets, the
relevant excitations are domain walls in the anti-ferromagnetic order.
In contrast, in the dimer model the relevant excitations correspond to
low lying magnons. In both cases these excitations experience
intraband transitions.  

An outline of the paper is as follows. In Section II, using first
order degenerate perturbation theory, we determine the dimer model's
low lying spectrum and the corresponding matrix elements. In Section
III we discuss in detail how to use these ingredients to compute the
susceptibility, $\chi^{zz} (\omega, Q)$. In particular, we explain the
use of the resummation implied by the Dyson-like equation. In the
final part of the paper, Section IV, we present the actual results for 
$\chi^{zz} (\omega, Q)$.
%%%%%%%%%%%%%%%%%%%%%%%%%%%%%%%%%%%%%%%%%%%%%%%%%%%%%%%%%%%%%%%%%%%%%%%
\section{Ground state and excited states of weakly coupled dimers}
\label{secstates}
%%%%%%%%%%%%%%%%%%%%%%%%%%%%%%%%%%%%%%%%%%%%%%%%%%%%%%%%%%%%%%%%%%%%%%%
Our starting point is the Hamiltonian given in Eqn. (\ref{eiii}) with
an even number of sites, $N$, and periodic boundary conditions. For
small $\alpha = J'/J$, we split the Hamiltonian into a solvable part
proportional to $J$, $H_0$, and a perturbation $H'$ (proportional to
$J'$):   
\begin{align}
H_{\rm dimer} = & \sum_{a=0}^{N/2-1}\left( J \mathbf{S}_{2a}
\cdot  \mathbf{S}_{2a+1} +J' \mathbf{S}_{2a+1} \cdot
\mathbf{S}_{2a+2} \right)  \nn
 = & H_0+H'.
\end{align}

We first make some remarks about the $J'=0$ case, in which the spins decouple into pairs on the bonds $J$. The ground state of $H_0$ is unique and is given by 
\begin{align}
& \ket{0} = \prod_{a=0}^{N/2-1} \ket{0}_{a}, \\
& \ket{0}_a = \frac{1}{\sqrt{2}} \left(\ket{\uparrow}_{2a}\ket{\downarrow}_{2a+1}-\ket{\downarrow}_{2 a}\ket{\uparrow}_{2 a+1}\right)
\end{align}
so that $\ket{0}_0$ is a singlet between sites 0 and 1. We take the associated eigenvalue, $E_0=-3NJ/8$, as the zero of energy.
\subsection{Excitations}
Excitations are formed by breaking singlets to create triplets. The
spectrum of $H_0$ then consists of degenerate levels at energies $nJ$
relative to the ground state, where $n<N/2$ is the number of
triplets. These excitations are dispersionless hardcore bosons. When the
perturbation $H'$ is applied the degeneracies are removed and coherent
single-particle excitations with dispersion relation $\epsilon_p$ are
formed. These magnons are not free, but interact with each other
through the perturbation $H'$ in addition to being subject to the
hard-core constraint. The first excited state consists of $N/2-1$
singlets and one triplet, leading to a total spin $S_{\rm tot}=1$.  

We define $d_{a}(m)$ as the operator that breaks a dimer between sites $2a$ and $2a+1$, 
creating a state with $z$-component of spin $m$.
For example the explicit form of one of these operators is
\begin{align}
d_{a}(0) = d_{a}^\dagger(0) = 2 S^z_{2a} \label{dimerop}
\end{align}
though it is important to realise that $d(\pm1) \ne d^\dagger(\pm1)$.
A translationally invariant state is formed by taking the Fourier transform:
\begin{align}
\ket{p,m} & =\sqrt{\frac{2}{N}} \sum_{a=0}^{N/2-1} e^{2ipa} d_a(m) \ket{0}.
\end{align}
Here the factor of two in the exponential accounts for the interdimer distance.
Periodic boundary conditions lead to the quantization condition
\[e^{ipN} = 1,\]
so that
\[p=\frac{2\pi n}{N}, \qquad n = 0,1,2, \hdots , N/2-1. \]
Strictly at the point $\alpha=0$ these single particle excitations are $N/2$-fold degenerate with a flat dispersion $\epsilon_p=J$. A finite value of $\alpha$ causes magnons to `hop'. To first order in $\alpha$ the dispersion is
\begin{align}
& \epsilon_p  = J-\frac{J'}{2}\cos(2p),
\end{align}
resulting in a gap
\begin{align}
\Delta=J-\frac{J'}{2}.
\end{align}
Using translational and spin-rotational invariance, 
we can express two-magnon states (to lowest order in $\alpha$) in the form
\begin{multline} \label{eqn2part} \ket{p_1,p_2,S,m} = \mathcal{N}_S(p_1,p_2) \sum_{a=1}^{N/2-1} \sum_{b=0}^{a-1} \psi^S_{a b}(p_1,p_2)\\
\times  \Phi^{S,m}_{a b} \ket{0}.
 \end{multline}
Here $S=0,1,2$ and the normalisation $\mathcal{N}$ will, in general, be dependent on the linear and angular momenta. Explicit expressions for the spin part $\Phi^{S,m}_{ab}$ are given in appendix \ref{seccg}.
The wavefunction is given by
\begin{align}
\psi^S_{a b}(p_1,p_2) = e^{2i(p_1a+p_2b)}+A_{p_1p_2}^{S}e^{2i(p_1b+p_2a)}.
\end{align}
Embodied in the non-trivial relative phase, $A^S_{p_1p_2}$, is the magnon-magnon interaction. For $\alpha=0$ the form of $A$ is unspecified because the magnons cannot hop onto the same site and as such do not interact. To lowest order in $\alpha$ the correct basis in degenerate perturbation theory is given by requiring
\begin{multline}
\label{eqndiagcondition}
\mathcal{P}_2H'\ket{p_1,p_2,S,m} \\ = -\frac{J'}{2}[\cos(2p_1)+\cos(2p_2)] \ket{p_1,p_2,S,m}
\end{multline}
where $\mathcal{P}_2$ is the projection operator onto the two-particle states.
When the triplets in the sum given in (\ref{eqn2part}) are well separated ($\left| a-b \right| > 1$) the condition (\ref{eqndiagcondition}) is trivially satisfied, independently of $A$. When the triplets are neighbouring ($\left| a-b \right| = 1$) we find
\begin{subequations}
\label{eqnphase}
\begin{align}
A^0_{p_1p_2} & = -\frac{1+e^{-2i(p_1+p_2)}-2e^{-2ip_2}}{1+e^{-2i(p_1+p_2)}-2e^{-2ip_1}}, \\
A^1_{p_1p_2} & = -\frac{1+e^{-2i(p_1+p_2)}-e^{-2ip_2}}{1+e^{-2i(p_1+p_2)}-e^{-2ip_1}}, \\
A^2_{p_1p_2} & = -\frac{1+e^{-2i(p_1+p_2)}+e^{-2ip_2}}{1+e^{-2i(p_1+p_2)}+e^{-2ip_1}}.
\end{align}
\end{subequations}
The magnons therefore experience both an infinite onsite repulsion and a nearest neighbour momentum and spin dependent interaction.
Periodic boundary conditions and the restriction on the sums lead to the conditions
\begin{align}
\label{eqnBAE}
A^S_{p_1p_2}=(-1)^Se^{ip_1 N} \mbox{  and  } A^S_{p_1p_2}e^{ip_2 N}=(-1)^S\ ,
\end{align}
with the implication that the quantisation of the two-particle
momenta depends on the total spin, $S$. These are in fact the Bethe
Ansatz equations for the spin-$\frac{1}{2}$ XXZ chain where the
sectors $S=0,1,2$ correspond to anisotropies
$\Delta=1,+\frac{1}{2},-\frac{1}{2}$ respectively (and antiperiodic
boundary conditions for $S=1$). Solving these
equations for finite $N$ to find $p_1$ and $p_2$ is a numerical task
which we outline in the appendix. We note here however that the
solutions may be complex, leading to bound states, in agreement with
results in the literature\cite{tennant2,uhrig,brooksharris}. For
notational convenience, we define a phase shift by 
\begin{align}
\delta_{p_1p_2}^S=-\frac{i}{2}\ln\big( A_{p_1p_2}^S \big).
\end{align}
For real $p_1,p_2$ the normalization of a two-particle state is given by
\begin{multline}
\label{eqn2partnormalization}
\mathcal{N}_S(p_1,p_2)= \left[ \frac{N}{2}\left( \frac{N}{2}-1 \right) \right.\\
- \left.\frac{N}{2}\frac{\cos(2\delta_{p_1p_2}^S)-\cos(2p_1-2p_2-2\delta_{p_1p_2}^S)}{1-\cos(2p_1-2p_2)} \right]^{-1/2}.
\end{multline}
We note that two-magnon states have the symmetry
\begin{align}
\ket{p_1,p_2,S,m}=e^{-2i\delta^S_{p_1p_2}}\ket{p_2,p_1,S,m}.
\end{align}
To avoid an overcomplete basis we make the restriction $p_1 > p_2$.

%%%%%%%%%%%%%%%%%%%%%%%%%%%%%
\subsection{Matrix Elements}
%%%%%%%%%%%%%%%%%%%%%%%%%%%%%
For small $\alpha$ the gap to excitations is of order $J$ and
states with $n$ magnons are suppressed by a factor $\exp(-\beta nJ)$
in the thermal trace (\ref{eii}). At low temperatures, $\beta J \gg
1$, we then make the approximation that we only have to consider
processes where at most two magnons couple to the spin
operators. Their contribution can be determined by considering matrix
elements involving at most two magnons. Due to the isotropy in spin
space of $H_{\rm   dimer}$ we require only the $\chi^{zz}$ component
of the dynamical susceptibility. In addition the Hamiltonian is
invariant under translations by two sites. When evaluating (\ref{eii})
it is then sufficient to consider matrix elements of the form $\langle l |
S^z_j|m\rangle$ where $j=0,1$ and $l,m$ correspond to states with
zero, one or two magnons. Calculating such elements is simple to
zeroeth order in perturbation theory and the results are summarized in
Tables \ref{tab1} and \ref{tab2}. Certain matrix elements are not given in the tables because they are identically zero. This can be seen by taking account of the fact that the operator $S^z_l$ commutes with the total $z$-component of spin, leading to the transition selection rule $\Delta S^z =0$. In addition some elements are zero by inspection of the states given in (\ref{eqnstates}).

We define the functions
\begin{multline}
\label{eqn2partU}
U_S(p,p_1,p_2)
=\frac{N}{2}\mathcal{N}_S(p_1,p_2)\exp\big\{-i\big(\delta^S_{p_1p_2}+S\frac{\pi}{2}\big)\big\}\\
\times\left[ \frac{\sin(p-p_1+\delta^S_{p_1p_2}+S\frac{\pi}{2})}{\sin(p-p_1)}+
  \frac{\sin(p-p_2-\delta^S_{p_1p_2}+S\frac{\pi}{2})}{\sin(p-p_2)}\right], 
\end{multline}
and
\begin{multline}
\label{eqn2partV}
V_{S'S}(p'_1,p'_2,p_1,p_2) =\Big(\frac{N}{2}\Big)^2
\mathcal{N}_S(p_1,p_2)\mathcal{N}_{S'}(p'_1,p'_2) \\
\times \exp\big\{i(\delta_{p_1p_2}^S-\delta_{p_1'p_2'}^{S'}+(S-S')\frac{\pi}{2})\big\}\\
\times
\Big[ \frac{\sin(p_1-p'_1-\delta^S_{p_1p_2}+\delta^{S'}_{p'_1p'_2}-(S-S')\frac{\pi}{2})}{\sin(p_1-p_1')} \\
+\;\frac{\sin(p_2-p'_2+\delta^S_{p_1p_2}-\delta^{S'}_{p'_1p'_2}-(S-S')\frac{\pi}{2})}{\sin(p_2-p_2')} \\
+\;\frac{\sin(p_1-p'_2-\delta^S_{p_1p_2}-\delta^{S'}_{p'_1p'_2}-(S-S')\frac{\pi}{2})}{\sin(p_1-p_2')} \\
+\;\frac{\sin(p_2-p'_1+\delta^S_{p_1p_2}+\delta^{S'}_{p'_1p'_2}-(S-S')\frac{\pi}{2})}{\sin(p_2-p_1')} \Big],
\end{multline}
which are useful when calculating matrix elements that involve two particle states.
\begin{table}[b]
\caption{Non-zero matrix elements of the interband type for $S^z_j$ acting at sites $j=0,1$.}
\label{tab1}
\begin{center}
\begin{tabular}{|| c || c ||}
\hline
$\bra{0} S^z_{j} \ket{p,m}$ & $ (-1)^j\sqrt{\frac{1}{2N}} \delta_{m,0}$ \\ [1ex]
$\bra{p_1,p_2,0,0}S^z_j \ket{p,0}$ & $(-1)^j\sqrt{\frac{2}{3N^3}} U_0(p,p_1,p_2)$ \\[1ex]
$\bra{p_1,p_2,2,0}S^z_j \ket{p,0}$ & $(-1)^{j+1}\sqrt{\frac{4}{3N^3}} U_2(p,p_1,p_2)$ \\[1ex]
$\bra{p_1,p_2,1,1}S^z_j \ket{p,1}$ & $(-1)^{j+1}\sqrt{\frac{1}{N^3}} U_1(p,p_1,p_2)$ \\[1ex]
$\bra{p_1,p_2,1,-1}S^z_j \ket{p,-1}$ &$(-1)^{j} \sqrt{\frac{1}{N^3}}  U_1(p,p_1,p_2)$\\[1ex]
$\bra{p_1,p_2,2,1}S^z_j \ket{p,1}$ &$(-1)^{j+1}\sqrt{\frac{1}{N^3}}  U_2(p,p_1,p_2)$\\[1ex]
$\bra{p_1,p_2,2,-1}S^z_j \ket{p,-1}$ &$(-1)^{j+1}\sqrt{\frac{1}{N^3}} U_2(p,p_1,p_2)$ \\[1ex]
\hline
\end{tabular}
\end{center}
\end{table}
\begin{table}[t]
\caption{Non-zero matrix elements of the intraband type for $S^z_j$ acting at sites $j=0,1$.}
\label{tab2}
\begin{center}
\begin{tabular}{|| c || c ||}
\hline
$\bra{p',m}S^z_{j}\ket{p,m}$& $\frac{1}{N} (\delta_{m,1}-\delta_{m,-1})$ \\[1ex]
$\bra{p'_1,p'_2,1,0}S^z_j\ket{p_1,p_2,0,0} $ & $\sqrt{\frac{8}{3N^4}}V_{10}(p_1',p_2',p_1,p_2)$ \\[1ex]
$\bra{p'_1,p'_2,2,0}S^z_j \ket{p_1,p_2,1,0}$ & $\sqrt{\frac{4}{3N^4}}V_{21}(p_1',p_2',p_1,p_2)$ \\[1ex]
$\bra{p'_1,p'_2,1,\pm 1}S^z_j \ket{p_1,p_2,1,\pm 1}$ & $\mp\sqrt{\frac{1}{N^4}}V_{11}(p_1',p_2',p_1,p_2)$ \\[1ex]
$\bra{p'_1,p'_2,2,\pm 1}S^z_j \ket{p_1,p_2,1,\pm 1}$ & $\pm\sqrt{\frac{1}{N^4}}V_{21}(p_1',p_2',p_1,p_2)$\\[1ex]
$\bra{p'_1,p'_2,2,\pm 1}S^z_j \ket{p_1,p_2,2,\pm 1}$ & $\pm\sqrt{\frac{1}{N^4}}V_{22}(p_1',p_2',p_1,p_2)$\\[1ex]
$\bra{p'_1,p'_2,2,\pm 2}S^z_j \ket{p_1,p_2,2,\pm 2}$ & $\mp\sqrt{\frac{4}{N^4}}V_{22}(p_1',p_2',p_1,p_2)$\\[1ex]
\hline
\end{tabular}
\end{center}
\end{table}

%%%%%%%%%%%%%%%%%%%%%%%%%%%%%%%%%%%%%%%%%%%%%%%%%
\section{Spectral Representation and Resummation}
\label{secresum}
%%%%%%%%%%%%%%%%%%%%%%%%%%%%%%%%%%%%%%%%%%%%%%%%%
Taking the definition of the susceptibility in the Matsubara formalism
(\ref{eii}) it is helpful to expand in terms of operators at
even and odd sites: 
\begin{multline}
\chi^{zz}(\omega, Q)=-\int_0^\beta d\tau\ e^{i\omega_n\tau}
\frac{1}{N}\sum_{l,l'=0}^{N/2-1}e^{-i2Q(l-l')}
\\\times\Bigl[
\langle S^z_{2l}(\tau)\ S^z_{2l'}\rangle
+\langle S^z_{2l+1}(\tau)\ S^z_{2l'}\rangle e^{-iQ}
\\
+e^{iQ} \langle S^z_{2l}(\tau)\ S^z_{2l'+1}\rangle
+\langle S^z_{2l+1}(\tau)\ S^z_{2l'+1}\rangle\Bigr]
\biggr|_{\omega_n\rightarrow \eta-i\omega}.
\label{chi}
\end{multline}
Using translational symmetry and grouping terms according to magnon number, the susceptibility can be written as
\begin{align}
\label{eqnchiexpansion}
&\chi^{zz}(\omega, Q)\equiv\frac{1}{Z}\sum_{r,s=0}^\infty C_{rs},\nn
\begin{split}
&C_{rs}=-\int_0^\beta d\tau\ e^{i\omega_n\tau}
\frac{1}{N}\sum_{l,l'=0}^{N/2-1}e^{-i2Q(l-l')}\sum_{\gamma_r,\gamma_s}e^{-\beta
  E_{\gamma_r}}\\
&\qquad\times\ e^{-\tau[E_{\gamma_s}- E_{\gamma_r}]}\
e^{i2(l-l') [P_{\gamma_s}-P_{\gamma_r}]}
M_{\gamma_r\gamma_s}\biggr|_{\omega_n\rightarrow \eta-i\omega}.
\end{split}
\end{align}
Here ${\gamma_s}$ is a multi-index enumerating all $s$-particle states, 
$E_{\gamma_s}$ and $P_{\gamma_s}$ are the energy and momentum of the
excited state $|{\gamma_s}\rangle$, $Z$ is the partition function and 
\begin{align}
M_{{\gamma_r}{\gamma_s}}=&
|\langle {\gamma_r}|S^z_0|{\gamma_s}\rangle|^2
+e^{iQ}\langle {\gamma_r}|S^z_0|{\gamma_s}\rangle
\langle {\gamma_s}|S^z_1|{\gamma_r}\rangle\nn
&+|\langle {\gamma_r}|S^z_1|{\gamma_s}\rangle|^2
+e^{-iQ}\langle {\gamma_r}|S^z_1|{\gamma_s}\rangle
\langle {\gamma_s}|S^z_0|{\gamma_r}\rangle.
\end{align}
We have also suppressed the energy and momentum labels on $C_{rs}$ for notational simplicity.
Carrying out the Fourier transform we have
\bea
\label{eqnCrs}
C_{rs}=\sum_{{\gamma_r},{\gamma_s}} \frac{N}{4}\delta_{Q+P_{\gamma_r},P_{\gamma_s}}\ 
\frac{e^{-\beta E_{\gamma_r}}-e^{-\beta E_{\gamma_s}}}
{\omega+i\eta+E_{\gamma_r}-E_{\gamma_s}}M_{{\gamma_r}{\gamma_s}}.
\eea
The non-vanishing contribution at $T=0$ is obtained from $C_{10}+C_{01}$:
\begin{align}
\begin{split}
C_{10}+C_{01}&=\frac{1-\cos(Q)}{4}\left(1-e^{-\beta\epsilon_Q}\right) \\
&\qquad\times \left[
\frac{1}{\omega+i\eta-\epsilon_Q}-
\frac{1}{\omega+i\eta+\epsilon_Q}\right]
\end{split}\\
&\equiv\left(1-e^{-\beta\epsilon_Q}\right) D(\omega,Q).
\end{align}
$D(\omega,Q)$ is then the bare magnon propagator. The remaining $C_{rs}$ terms, up to $C_{12}+C_{21}$, are obtained in a simple manner using Eqn. (\ref{eqnCrs}) and the matrix elements in Tables \ref{tab1} and \ref{tab2}:
\be
C_{11} =\frac{1+\cos(Q)}{N} \sum_p \frac{e^{-\beta \epsilon_p}-e^{-\beta\epsilon_{p+Q}}}{\omega+\epsilon_p-\epsilon_{p+Q}+i\eta}, \label{eqnintra}
\ee
\bea
&&C_{12}+C_{21} =\frac{1-\cos(Q)}{4}\left(\frac{2}{N}\right)^2 \sum_S
\sum_{p_1>p_2}\left(\frac{2S+1}{3}\right) \nn 
&& \times U_S^2(Q+p_1+p_2,p_1,p_2)\!\left[e^{-\beta \epsilon_{Q+p_1+p_2}}-e^{-\beta
  (\epsilon_{p_1}+\epsilon_{p_2})}\right] \nn 
&& \times
\left\{\frac{1}{\omega+\epsilon_{Q+p_1+p_2}-\epsilon_{p_1}-\epsilon_{p_2}+i\eta}
\right. \nn
&&\qquad\qquad-\left.\frac{1}{\omega-\epsilon_{Q+p_1+p_2}+\epsilon_{p_1}+\epsilon_{p_2}+i\eta}
\right\}. 
\label{eqn1221}
\eea
We reiterate that the allowed values of $p_1,p_2$ in the sum above depend on $S$ and that the sum is only over those momenta that produce unique two particle states.

We now make some initial remarks about the structure of the $C_{rs}$
terms. Firstly, the two site dimer basis leads to a $Q$ dependent
prefactor that differs between the interband terms
$C_{r,r+1},C_{r+1,r}$ and intraband terms $C_{rr}$.  Secondly, the
$C_{12}+C_{21}$ term will diverge with system size as $N$. This
divergence is expected and should cancel with terms arising from an
expansion of the partition function $Z$. Lastly, the $\omega$
dependence in the denominators is such 
that interband terms diverge as $\omega\to\epsilon_Q$. This divergence
is reflected in the intraband terms, which diverge for $\omega \to \pm
J'\sin(Q)$. Higher order terms have stronger divergences. This
standard behaviour is a consequence of the essentially nonperturbative
nature of the finite $T$ magnon lifetime. Perturbation theory is
unable to capture, order-by-order, the decay enhancing effect of
particle-particle scattering processes, hence lifetimes remain
infinite to all orders. Treating the interaction accurately and
rendering the lifetimes finite requires a summation of perturbation
theory terms to infinite order. We achieve this infinite summation for
certain processes as follows. Taking into account interactions between
the magnons by a Dyson-like equation, we conclude that we can write  
\bea
\chi^{zz}(\omega,Q)&=& \frac{D(\omega,Q)}{1-D(\omega,Q)\Sigma(\omega,Q)}.
\label{selfenergy}
\eea
Expanding this we obtain
\bea
\chi^{zz}(\omega,Q)= \left[D(\omega,Q)+D^2(\omega,Q)\Sigma(\omega,Q)+\ldots \right]
\eea
On the other hand, from our low temperature expansion of the spectral representation we have
\begin{align}
\chi^{zz}&(\omega,Q)=\frac{1}{Z}
\left[C_{01}+C_{10}+C_{11}+C_{12}+C_{21}+\ldots\right]\nn
=&\frac{1}{1+Z_1+Z_2+\ldots} \nn
& \Big[\left(1-e^{-\beta\epsilon_Q}\right) D(\omega,Q)+C_{11}+C_{12}+C_{21}+\ldots\Big]\nn
=&\left(1-e^{-\beta\epsilon_Q}\right) D(\omega,Q)+
C_{11}+C_{12}+C_{21}\nn
& \quad - Z_1\left(1-e^{-\beta\epsilon_Q}\right)
D(\omega,Q)+ \ldots\nn
=& D(\omega,Q)+\Big[(C_{11}+C_{12}+C_{21})\nn
& \quad -Z_1\left(1-e^{-\beta\epsilon_Q}\right)
D(\omega,Q)-e^{-\beta\epsilon_Q}D(\omega,Q)\Big] + \hdots
\end{align}
The contribution to the partition function from the one particle states, $Z_1=3\sum_p e^{-\beta\epsilon_p}$, must be included to cancel the $N$ dependence of the $C_{12}+C_{21}$ contribution. Comparing the two expansions, we make the identification
\begin{align}
\label{eqnselfen}
\Sigma(\omega,Q)\approx & D^{-2}(\omega,Q)\Big[
(C_{11}+C_{12}+C_{21}) \nn & \quad -\left(Z_1\left(1-e^{-\beta\epsilon_Q}\right)
+e^{-\beta\epsilon_Q}\right)D(\omega,Q)\Big].
\end{align}
Finally we calculate the quantity of experimental interest as 
\begin{multline}
\label{eqndsfresummed}
S^{zz}(\omega, Q)=-\lim_{\eta \to 0}\frac{1}{\pi}\frac{1}{1-e^{-\beta\omega}}
\\{\rm Im}\left[
\frac{D(\omega,Q)}{1-D(\omega,Q)\Sigma(\omega,Q)}\right].
\end{multline}
%%%%%%%%%%%%%%%%%%%%%%%%%%%%%%%%%%%%%
\section{Results and Discussion}
%%%%%%%%%%%%%%%%%%%%%%%%%%%%%%%%%%%%%
We now choose $J=1$ and $J'=0.1$ so that $\alpha=0.1$ is small, as
required by our expansion. We consider temperatures less than the gap,
so that the low magnon number approximation holds. We then calculate
the dynamical structure factor at given $\omega$ and $Q$
numerically. Sums over momenta such as those in Eqs. (\ref{eqnintra})
and (\ref{eqn1221}) are performed for systems of $N/2$
dimers. Analytically, the standard procedure for evaluating
(\ref{eii}) is to take the thermodynamic limit, perform the resulting
momentum integrals, extract the imaginary part of the resulting
susceptibility and finally take the limit $\eta \to 0$. The zero
temperature result for $S^{zz}(\omega,Q)$ then takes the form of a
delta function at $\omega=\epsilon_Q$. For the purposes of numerics,
it is necessary to stipulate $\eta$ before performing the sums. This
results in a broadened zero temperature result, a Lorentzian peak of
width $\eta$. To obtain accurate results at finite $T$, this width
must be small in comparison with the thermally activated broadening
which scales as  $J'e^{-\beta J}$. In contrast, to avoid finite size
effects, the sums must be evaluated on a grid in wave vector space
that is fine enough to resolve the Lorentzian. The condition for
producing accurate numerics is then 
\begin{align}
e^{-\beta J} \gg \frac{2\eta}{J'} > \frac{4 \pi}{N}.
\end{align}
In principle by increasing the system size, $N$, very small values of $\eta$ could be used. However as explained above this is only necessary at very low temperatures, where the broadening is not sufficiently asymmetric to be interesting.
For the range of temperatures we investigate we find a suitable value of $\eta$ to be 0.002. When calculating the intraband response we typically use systems of size $N/2=600$. The
interband response is less sensitive to finite size effects and sums
using a smaller system size, $N/2=400$, are permissible. We find that
the effects of using a larger number of dimers are negligible. If
the delta function in Eqn.(\ref{eqnCrs}) is ever to be satisfied we
must restrict the external momentum to $Q=4\pi n/N$ with integer
$n$. We leave discussion of further issues affecting numerical
accuracy, in particular bound state solutions of Eqns. (\ref{eqnBAE})
to the appendix. 

\begin{figure}[b]
\begin{center}
\epsfxsize=0.48\textwidth
\epsfbox{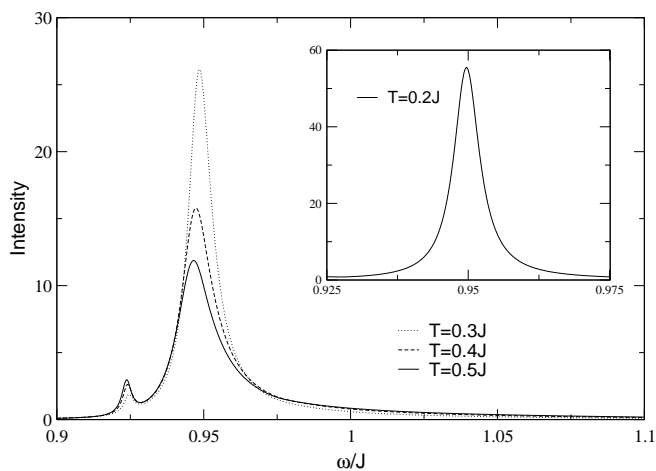}
\caption{The one-magnon response at $Q=\pi$ for $N/2=400$. At $T=0.2J$ the lineshape is nearly Lorentzian (inset), but is increasingly asymmetric as temperature approaches the energy gap.}
\label{figtemps}
\end{center}
\end{figure}

We first consider the behaviour of the one-magnon or interband
response and examine the lineshape in energy by fixing the external
wavevector $Q$. Fig. \ref{figtemps} shows the dynamical structure
factor at $Q=\pi$ for a range of temperatures. At temperatures less
than the gap the main feature is a peak at $\omega=\epsilon_Q$. At
temperatures below $T\sim0.3J$ the peak is approximately
Lorentzian. The maximum response falls rapidly with increasing
temperature and the peak broadens asymmetrically. In particular the
peak becomes skewed, with a tail extending towards energies
$\omega\sim J$.

The degree of asymmetry is large compared to that found for a variety
of other spin chains using a semiclassical approach\cite{young,damle}.
Instead it is similar to the asymmetry found by exact
diagonalization \cite{mikeska}. The broadening also resembles that
found for two integrable spin chains in Ref. [\onlinecite{essrmk}]
which uses the same resummation scheme as this paper. Most importantly
the asymmetry has been found to be consistent with recent data on
copper nitrate\cite{tennant1}. We also point out that the asymmetry is
primarily a consequence of the two-magnon interband terms
$C_{12}+C_{21}$. Intraband terms, $C_{rr}$, have vanishing spectral
weight in this region and their lack of influence is confirmed by
observing that the one-magnon mode is unaffected if we neglect them
in the resummation.

\begin{figure}[t]
\begin{center}
\epsfxsize=0.48\textwidth
\epsfbox{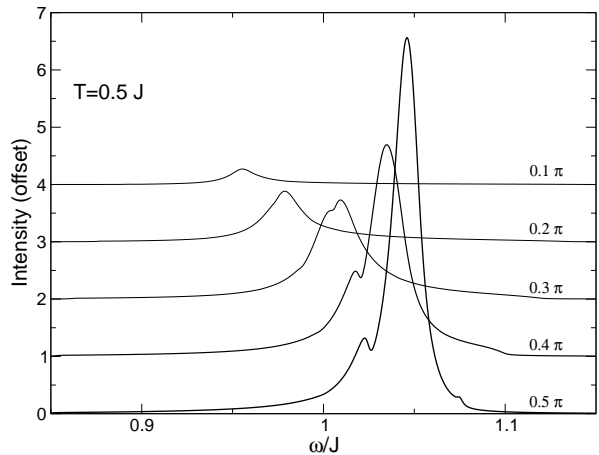}
\caption{Wavevector dependence of the one-magnon or interband
response. The peak positon is given by $\omega=\epsilon_Q$.  At
$Q=0$ the response vanishes. Top: Wavevectors between $Q=2\pi/10$
and $\pi/2$. The vertical axis is offset by integer values $5-m$ for
$Q=m\pi/10$.}
\label{figQscan1a}
\end{center}
\end{figure}

\begin{figure}[t]
\begin{center}
\epsfxsize=0.48\textwidth
\epsfbox{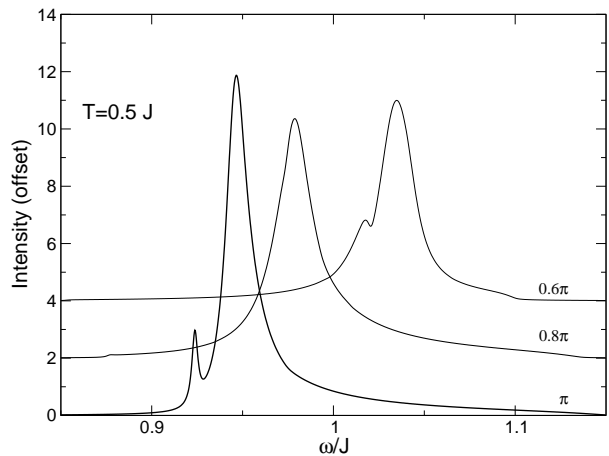}
\caption{Wavevector dependence of the one-magnon or interband
response. The peak positon is given by $\omega=\epsilon_Q$.  At
$Q=0$ the response vanishes. Wavevectors between $Q=3\pi/5$ and
$\pi$. The vertical axis is offset by integer values $10-m$ for
$Q=m\pi /10$.}  
\label{figQscan1b}
\end{center}
\end{figure}

The small peak at lower energies seen in Figs. \ref{figtemps},
\ref{figQscan1a} and \ref{figQscan1b} is due to processes that involve
bound states. On the level of the calculation presented in this paper,
the bound state response is essentially sharp and grows with
temperature. This behaviour is an artefact of the order at which we
truncate the perturbative expansion. Taking into account higher order
corrections to the magnon dispersion would broaden the
peak. Similarly, taking into account processes involving e.g. a bound
state and a magnon in the thermal background and a bound state and two
magnons in the intermediate state would lead to a significant
weakening of this feature.

The behaviour in wavevector space is dominated by the static structure
factor, $1-\cos(Q)$, arising from the two site dimer basis. As a
result, the response is maximal for $Q=(2n+1)\pi$ and disappears
altogether at $Q=2n\pi$ as in Figs. \ref{figQscan1a} and \ref{figQscan1b}
(for integer $n$).  
Exactly at $Q=(2n+1)\pi/4$ the susceptibility, $\chi^{zz}$, is a
symmetric function of $\omega$ about the point $\omega=J$; however the
dynamical structure factor, $S^{zz}$, is not symmetric because
spectral weight is shifted to higher energies by the factor
$(1-e^{-\beta \omega})^{-1}$. 

The direction of asymmetry in the dynamical response at $\omega\sim
\epsilon_Q$ can be understood qualitatively in terms of a joint
density of states for transitions between occupied one magnon and
unoccupied two magnon states. For $\omega>0$ this takes the form  
\bea
N_{1\to 2} = \sum_{p,p_1,p_2} n(p)\bar{n}(p_1,p_2)
%\nn
%&&\quad\times\ 
\delta_{Q+p,p_1+p_2}
\delta_{\omega+\epsilon_p,\epsilon_{p_1}+\epsilon_{p_2}}.
\eea
Here $n(p)$ is the thermal occupation number for a
one-magnon state with momentum $p$ and $\bar{n}(p_1,p_2)$ is the
probability that the two-magnon state characterized by momenta
$p_1$ and $p_2$ is unoccupied. At low temperatures and weak
inter-dimer interactions we have approximately 
\be
n(p)\approx e^{-\beta \epsilon_p}\ ,\quad
\bar{n}(p_1,p_2)\approx (1-e^{-\beta\epsilon_{p_1}})(1-e^{-\beta\epsilon_{p_2}}).
\ee
For $-\pi/4 < Q < \pi/4$ and $\pi/4 < Q < 3\pi/4$ this function is
skewed towards higher and lower energies respectively. On the other
hand the specific form of the lineshape is dictated by the matrix
element $M_{12}$ and hence by the magnon-magnon interaction. 
The fact that the lineshape of the dynamical response at
$\omega\approx \epsilon_Q$ is skewed towards low frequencies in some
regions of the Brillouin zone and towards high frequencies in others
is a consequence of the smallness of the ratio of bandwidth to magnon gap.
This should be contrasted to the findings of Ref. [\onlinecite{essrmk}]
for the lineshape in the O(3) nonlinear sigma model, for which the
bandwidth is infinite and concomitantly the asymmetry was found to
always extend towards higher energies.

\begin{figure}[ht]
\begin{center}
\epsfxsize=0.48\textwidth
\epsfbox{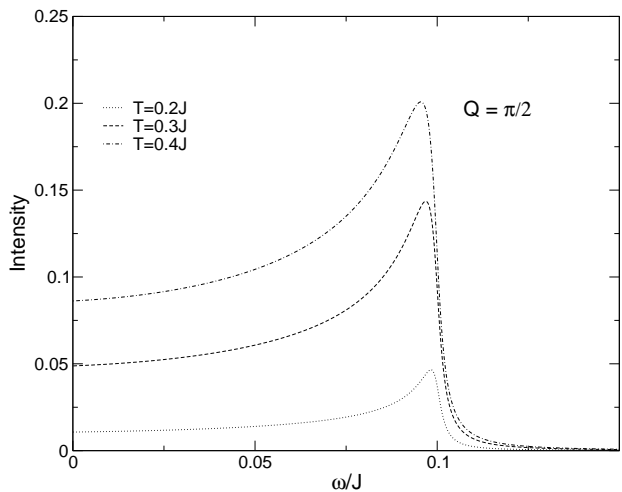}
\caption{Development of intraband scattering with temperature at
  $Q=\pi/2$ for $N/2=600$ dimers.} 
\label{figintra}
\end{center}
\end{figure}
Next we turn to the intraband response. When magnons are thermally
excited, incident neutrons can scatter them within the same band with
energy transfers small compared to the gap. Accordingly at finite
temperatures there is a spin response at energies $\omega \sim 0$. To
lowest order the contribution to the intraband response is given by  
\begin{multline}
-\frac{1}{\pi}\frac{1}{1-e^{-\beta \omega}} \mathrm{Im} C_{11} \\
= \frac{1+\cos(Q)}{2\pi} \frac{e^{-\beta(J-\omega/2)}}{\sqrt{(J' \sin(Q))^2-\omega^2}} \\
\times\cosh\Big(\frac{\beta}{2}\cot(Q)\sqrt{(J' \sin(Q))^2-\omega^2}\Big),
\end{multline}
valid for $\lvert \omega \rvert \le J'\lvert \sin(Q) \rvert$. This
response is bounded by inverse square root singularities and has an
overall magnitude that grows with temperature as $e^{-\beta J}$. We
have calculated the next leading contribution $C_{22}$ using the
matrix elements in Table \ref{tab2} and found it to exhibit a stronger
divergence at $\omega=\pm J'\sin(Q)$. This shows that, just as for
the interband contributions, a resummation needs to be carried
out. This is achieved by including the intraband scattering
contributions in the low temperature expansion of the self-energy in
Eqn. (\ref{selfenergy}). A complication that arises in doing so is that
because of the different prefactors $1\pm\cos(Q)$ for intra and
interband scattering the result of the resummation is reliable only at
very low temperatures for certain wave vectors. For such values of $Q$
higher order terms such as $C_{22}$ should be taken into
account. However, a consistent treatment of such terms would require
the incorporation of interband contributions involving three-magnon
states, which is beyond the scope of this work. By including $C_{11}$
in the resummation we remove the square root singularities and
associated threshold. Instead the response has two 
finite peaks and falls rapidly to zero for $\lvert\omega\rvert
>J'\lvert\sin(Q)\rvert$. This is physically sensible and is analogous
to what is found for the intraband scattering in the spin-1/2
Heisenberg-Ising chain \cite{JE}.  
In Fig. \ref{figintra} we show the calculated
intraband scattering at $Q=\pi/2$. The $Q$-dependent range in
$\omega$ of this scattering compares well with that calculated
previously by exact diagonalisation of chains of $N=16$
sites.\cite{mikeska} In that case however, the small system size
limited the number of available transitions and so the lineshape was
not representative of the thermodynamic limit.  
%\begin{figure}[t]
%\includegraphics[width=3in]{Tscan}
%\caption{Development of intraband scattering with temperature at $Q=\pi/2$ for $N/2=600$ dimers.}
%\label{figintra}
%\end{figure}

In conclusion we have calculated the approximate dynamical structure
factor at finite temperature of the alternating Heisenberg chain in
the limits of strong alternation and low temperature. The method we
use has previously been applied to integrable spin chains but in this
case the system is non-integrable. We find that the lineshape of the
lowest lying one-magnon mode is increasingly asymmetric with
temperature, a direct consequence of magnon-magnon interactions. We
also establish a prediction for the low temperature lineshape of the
intraband scattering.  

\begin{acknowledgments}
We are grateful to Bella Lake and Alan Tennant for numerous
enlightening discussions. This work was supported by the EPSRC under grant GR/R83712/01 (FHLE and AJAJ), the DOE under
contract DE-AC02-98 CH 10886 (RMK) and the ESF network INSTANS.
\end{acknowledgments}

\appendix
\section{Two-Magnon States}
\label{seccg}
The spin part of the two-magnon states is of the form
\begin{align}
\Phi_{ll'}^{S,m} & = \sum_{\{m_1,m_2\}} c_{m_1,m_2}^{S,m}d_l(m_1)d_{l'}(m_2).
\end{align}
The $c$'s are Clebsch-Gordan coefficients but the explicit expressions are given below for convenience:
\begin{subequations}
\label{eqnstates}
\begin{align}
\begin{split}
\Phi_{ll'}^{0,0} & = \frac{1}{\sqrt{3}} \Big[ d_l(1)d_{l'}(-1)+d_l(-1)d_{l'}(1)\\
&\qquad-d_l(0)d_{l'}(0) \Big],
\end{split} \\
\Phi_{ll'}^{1,0} & = \frac{1}{\sqrt{2}} \Big[ d_l(1)d_{l'}(-1)-d_l(-1)d_{l'}(1) \Big],
 \\
\begin{split}
\Phi_{ll'}^{2,0} & = \frac{1}{\sqrt{6}}\Big[ d_l(1)d_{l'}(-1)+d_l(-1)d_{l'}(1)\\
&\qquad+2d_l(0)d_{l'}(0) \Big] ,
\end{split}\\
\Phi_{ll'}^{1,1} & = \frac{1}{\sqrt{2}}\Big[ d_l(1)d_{l'}(0)-d_l(0)d_{l'}(1) \Big], \\
\Phi_{ll'}^{1,-1} & = \frac{1}{\sqrt{2}} \Big[ d_l(0)d_{l'}(-1)-d_l(-1)d_{l'}(0) \Big],
\\
\Phi_{ll'}^{2,1} & = \frac{1}{\sqrt{2}}\Big[ d_l(1)d_{l'}(0)+d_l(0)d_{l'}(1) \Big], \\
\Phi_{ll'}^{2,-1} & = \frac{1}{\sqrt{2}}\Big[ d_l(0)d_{l'}(-1)+d_l(-1)d_{l'}(0) \Big],\\
\Phi_{ll'}^{2,2} & = d_l(1)d_{l'}(1), \\
\Phi_{ll'}^{2,-2} & = d_l(-1)d_{l'}(-1).
\end{align}
\end{subequations}

%%%%%%%%%%%%%%%%%%%%%%%%%%%%%%%%%%%%%%%%%%%%%%%%%%%%%%%%%%%%%%
\section{Quantization of the Two-Magnon Momenta}
%%%%%%%%%%%%%%%%%%%%%%%%%%%%%%%%%%%%%%%%%%%%%%%%%%%%%%%%%%%%%%
\label{secmomenta}
In order to carry out momentum sums over two-particle states on a
finite lattice we  require knowledge of the allowed values of the
momenta, $p_1$ and $p_2$ in each sector, $S$. In practice, this means
we must solve equations (\ref{eqnphase}) and (\ref{eqnBAE})
numerically to find the $(N/2-1)N/4$ pairs $\{p_1,p_2\}$ allowed by
the condition $p_1 > p_2$. This is a problem usually
encountered in models solvable by Bethe ansatz \cite{book}.
%%%%%%%%%%%%%%%%%%%%%%%%%%%%%
\subsection{Real Solutions}
%%%%%%%%%%%%%%%%%%%%%%%%%%%%%
\label{subsecreal}
We first consider scattering states of two magnons, for which $p_1$
and $p_2$ are both real. In each spin sector $S$ the (XXZ Bethe
ansatz) equations (\ref{eqnBAE}) can be written in the form 
\bea
\label{eqnuseless}
e^{iNp_1}  =(-1)^S A_{p_1p_2}^{S}\ ,\quad
e^{iNp_2}  =(-1)^S A_{p_2p_1}^{S}.
\eea
In order to enumerate all roots of the coupled equations
(\ref{eqnuseless}) we take the logarithm. We choose a branch cut such
that 
\begin{align}
p_1 & =-\frac{i}{N}\ln\big(- A_{p_1p_2}^{S}  \big)+\frac{\pi}{N}
(2I_1+1) \ ,\nn
p_2 & =-\frac{i}{N}\ln\big(- A_{p_2p_1}^{S}  \big)+\frac{\pi}{N}
(2I_2+1),
\label{eqnuseful1}
\end{align}
for $S=0,2$ and
\begin{align}
p_1 & =-\frac{i}{N}\ln\big(- A_{p_1p_2}^{1}  \big)+\frac{\pi}{N}
2I_1 \ ,\nn
p_2 & =-\frac{i}{N}\ln\big(- A_{p_2p_1}^{1}  \big)+\frac{\pi}{N}
2I_2
\label{eqnuseful2}
\end{align}
for S=1.
Here the integers $I_{1,2}$ have range $0\leq I_{1,2}<N/2$.
We note that $I_1\geq I_2$ implies that $p_1>p_2$ and using the
indistinguishability of particles we can restrict ourselves without
loss of generality to the case $I_1>I_2$. Using the parametrization
(\ref{eqnuseful1},\ref{eqnuseful2}) it is now a relatively straightforward matter to
determine real roots corresponding to pairs of integers $I_1>I_2$ by
standard numerical root finding algorithms.

There are a number of roots which require special
treatment. Specifically in the singlet sector, there is a class of
real roots $\{p_1=2(I_1+1)\pi/N$, $p_2=0\}$ and in the
quintet sector there is a solution $\{p_1=\pi/2,p_2=0\}$ (for a system with $N/2$ even). For these
cases the derivation of equations
(\ref{eqn2partnormalization},\ref{eqn2partU},\ref{eqn2partV}) needs to
be revisited. For the special solutions in the singlet sector
the phase shift is zero and one finds (setting $p_1=q$)
\begin{align}
\mathcal{N}^2(q,0) &= \Big[\Big(\frac{N}{2}\Big)^2-N \Big]^{-1}\\
U(p,q,0) &= \mathcal{N}(q,0) \left(2-\frac{N}{2}
\delta_{p,0}-\frac{N}{2}\delta_{p-q,0} \right). 
\end{align}
In the quintet sector the same expression is found, but
the arguments of the two Kronecker deltas above are never
satisfied. 

The matrix elements for $C_{22}$ involving the special solutions are
similarly affected and need to be replaced by
\bea
&&V_{S'S}(q',0,p_1,p_2) = -\mathcal{N}(q',0)\mathcal{N}_S(p_1,p_2)\Big(\frac{N}{2}\Big)^2
e^{i\delta_{p_1p_2}^S}\nn
&&\qquad\times \ \Big[2\cos(\delta_{p_1p_2}^S)
+\frac{\sin(p_1-q'-\delta_{p_1p_2}^S)}{\sin(p_1-q')}\nn
&&\quad\qquad\qquad+\frac{\sin(p_2-q'+\delta_{p_1p_2}^S)}{\sin(p_2-q')} \Big],
\eea
\be
V_{S'S}(q',0,q,0) =
\mathcal{N}(q',0)\mathcal{N}(q,0)\Big(\frac{N}{2}\Big)^2 \Big[
  \frac{N}{2}-4 \Big]\ ,
\ee
\bea
&&V_{SS}(p_1,p_2,p_1,p_2)  =
2\mathcal{N}^2_S(p_1,p_2)\Big(\frac{N}{2}\Big)^2\nn
&&\qquad\qquad\times\ \Big[\frac{N}{2}-1
-\frac{\sin(p_1-p_2-2\delta_{p_1p_2}^S)}{\sin(p_1-p_2)} \Big].
\eea
The numerical root finder does not converge for ${\cal O}(N)$ pairs
of integers $\{I_1^c,I_2^c\}$. Most of these correspond to complex
solutions of (\ref{eqnuseful1},\ref{eqnuseful2}), which are discussed in the next
subsection. 
So far we have restricted our discussion to real roots with distinct
integers $I_1\neq I_2$. The reason for this restriction is that
$I_1=I_2$ correponds generically to $p_1=p_2$, which does not yield a
valid solution of the Schr\"odinger equation. However, in analogy to
what was shown in Ref. [\onlinecite{essxxx}], there are additional ``good''
real solutions with repeating integers $I_1=I_2$, which have to be
treated with care.

%%%%%%%%%%%%%%%%%%%%%%%%%%%%%%%
\subsection{Complex Solutions}
%%%%%%%%%%%%%%%%%%%%%%%%%%%%%%%
\label{seccmplxsol}
In addition to real roots there exist complex solutions of
(\ref{eqnuseful1},\ref{eqnuseful2}). These give rise to wave functions that exhibit an
exponential decay with respect to the distance between the two magnons
and hence correspond to bound states. As the equations (\ref{eqnBAE})
are closed under complex conjugation, complex roots must come in pairs
\be
p_1=x+iy\ ,\qquad p_2=x-iy\ ,\ x,y\ \text{real}.
\ee
Adding the two momenta using (\ref{eqnuseful1},\ref{eqnuseful2}) gives
\begin{align}
p_1+p_2 & = 2x = \pi \frac{2}{N} (I^c_1+I^c_2+1),\quad& S=0,2,\nn
p_1+p_2 & = 2x = \pi \frac{2}{N} (I^c_1+I^c_2),  \quad& S=1.
\end{align}
The real component $x$ is now uniquely defined by the integers $I_1^c,I_2^c$. Defining $\phi_s=S(S+1)/2-2$ and substituting for $x$, the value of $y$ can be found by solving
\begin{gather}
 e^{ip_1N}=(-1)^S A_{p_1p_1^\star}^S \nn
\intertext{or equivalently}
e^{ixN}e^{-yN}+(-1)^S\frac{1+e^{-4ix}+\phi_s e^{-2ix}e^{-2y}}{1+e^{-4ix}+\phi_s e^{-2ix}e^{2y}}=0
\end{gather}
using a Newton-Raphson method. The resulting values of $y$ are such that $A_{p_1p_1^\star}^S$ or equivalently $e^{ip_1N}$ is either very small or very large. Consequently, in order to demonstrate that our solutions satisfy both Bethe Ansatz equations (\ref{eqnBAE}) numerically, the value of $p_1$ must be known to very high precision. Fortunately the matrix elements are less sensitive and for the system sizes we consider it transpires that to evaluate $C_{12}+C_{21}$ accurately $17$ significant figures of $p_1$ are sufficient.
The two-particle normalization for complex momenta is given by
\begin{multline}
\mathcal{N}_S(x+iy,x-iy)  = e^{yN/2}\Big[(-1)^S e^{ixN}\frac{N}{2}\left( \frac{N}{2}-1 \right)\\- \frac{N}{2}\frac{\cosh(yN)-\cosh(4y-Ny)}{1-\cosh(4y)} \Big]^{-1/2},
\end{multline}
and for the matrix elements the modification is
\begin{multline}
U_S(p,p_1,p_1^\star) = \frac{N}{2} \mathcal{N}_S(x+iy,x-iy)e^{i\frac{\pi}{2}S} \\
\Big\{\left(1+A_{p_1p_1^\star}^S \right)\cos(2(p-x)-\frac{\pi}{2}S)\nn
-\left(e^{-2y}+A_{p_1p_1^\star}^Se^{2y} \right)\cos(\frac{\pi}{2}S)\Big\}\nn
\big(\cos(2p-2x)-\cosh(2y)\big)^{-1},
\end{multline}
which is an even function of $y$, as required.

\end{document}